\documentstyle[rotate,epsf]{mn}

\newcommand{\Msun}{\mbox{M$_{\odot}$}}
\newcommand{\chis}{\mbox{$\chi^{2}~$}}
\newcommand{\rchis}{\mbox{$\chi^{2}_{\nu}~$}}
\newcommand{\kms}{\mbox{\,km\,s$^{-1}$}}

\newcommand{\ergss}{\mbox{\,erg\,s$^{-1}$}}

\title [The outburst radial velocity curve of X-Ray Nova Scorpii 1994 
(=GRO J1655--40)]
{The outburst radial velocity curve of X-Ray Nova Scorpii 1994 (=GRO~J1655--40):
a reduced mass for the black hole?}
\author[S.N.~Phillips, T.~Shahbaz and Ph.~Podsiadlowski]{
S.N.~Phillips, T.~Shahbaz and Ph.~Podsiadlowski \\
University of Oxford, Department of Astrophysics, Nuclear Physics
Building, Keble Road, Oxford, OX1~3RH, UK }

\begin{document}

\maketitle

\begin{abstract}

\noindent
We present a reanalysis of the outburst radial velocity data for X-Ray Nova
Scorpii 1994. Using a model based on X-ray heating of the secondary star
we suggest a more realistic treatment of the radial velocity data.
Solutions are obtained in the ($K_{2},q$) plane which, when combined with
the published value for the binary mass ratio and inclination, constrain
the mass of the black hole to within the region $4.1<M_{1}<6.6$ \Msun\
(90 per cent confidence), which is significantly lower than the value
obtained by Orosz \& Bailyn (1997). This reduced lower bound for the
black hole mass together with the high space velocity of the system is
consistent with the idea that it was formed by the post-supernova
collapse of a neutron star.
\end{abstract}

\begin{keywords}
accretion, accretion discs -- binaries: close -- stars: individual: X-Ray Nova
Sco 1994 (GRO~J1655--40) -- X-rays: stars.
\end{keywords}

\section{Introduction}

X-ray novae are low-mass binary systems in which a compact object
undergoes unstable accretion from a late-type companion star, releasing
energy in the form of X-rays. In a number
of cases there is substantial evidence that the compact object is a black
hole (see Tanaka \& Shibazaki 1996 and van Paradijs \& McClintock 1995
for reviews). Some of the best evidence for the presence of a black hole
is obtained from the measurement of the orbital velocity of the companion
star, leading to a determination of the mass function of the system and
hence the minimum mass of the compact object. When this mass exceeds the
maximum mass for a neutron star ($\sim 3.2$ $\rm{M}_\odot$; Rhoades \&
Ruffini 1974), a black hole seems the only remaining possibility.

The X-ray nova GRO~J1655--40 is one such system, and was discovered on
July 27 1994 with BATSE on board the Compton Gamma Ray Observatory (Zhang
et al.\ 1994). It has been studied extensively during the past two years
in X-rays and at optical and radio wavelengths (Bailyn et al.\ 1995a and
b, Zhang et al.\ 1995, van der Hooft et al.\ 1998). Strong evidence that
the compact object in X-Ray Nova Sco is a black hole was presented by Bailyn et
al.\ (1995b) who initially established a spectroscopic period of 
$2.601\pm0.027$ days, and suggested a mass function 
$f(M)=3.16\pm0.15$ \Msun. An improved value of
$f(M)=3.24\pm0.09$ \Msun\ was presented by Orosz \& Bailyn
(1997), derived from a radial velocity semi-amplitude $K_{2}=228.2\pm2.2$
\kms. Their fitted values of inclination and the mass
ratio then implied a black hole mass of $M_1=7.01\pm0.22$ \Msun.

However, in calculating the radial velocity semi-amplitude, from which
the mass function is derived, Orosz \& Bailyn (1997) used both quiescent
data (taken in 1996 February 24--25) \emph{and} outburst data (taken in
1995 April 30--May 4), while Bailyn et al. (1996b) used just outburst
data, and in both cases a sinusoidal fit was performed. We suggest that
using outburst data in this way may lead to an incorrect result. The
effect of substantial heating of the secondary can shift the `effective
centre' of the secondary, weighted by the strength of the absorption
lines, from the centre of mass of the star, as described in section 2.
This results in a significant distortion of the radial velocity curve
and renders a sinusoidal fit to be clearly inadequate, leading to a
spuriously high radial velocity semi-amplitude. The masses of the
binary components derived from this will therefore be incorrect.

In this paper, we intend to consider only the outburst radial velocity
data (from 1995 April 30--May 4). We use a model which incorporates the
basic effects of X-ray heating of the secondary, and attempt to derive a more
realistic range for the radial velocity semi-amplitude. Using the orbital
period and the range of inclinations obtained by van der Hooft et al.\ (1998), 
we obtain ($K_{2},q$) solutions. Using these solutions in conjunction with
the mass ratio (Orosz \& Bailyn 1997), we determine new limits on the
masses of the secondary star and the black hole. Finally, we consider 
the implications of this on some current evolutionary scenarios for the 
black hole in X-Ray Nova Scorpii 1994.

\section{The radial velocity curve}

It is generally believed that any initial non-circularity in the orbit of
the binary system would have been rapidly removed by tidal forces between
the secondary star and the black hole, and that the present orbits are
indeed circular. However, Davey and Smith (1992) argue that the radial
velocity curves may still be distorted from a pure sine wave by
geometrical distortion and heating of the secondary star by the compact
object, causing the centre of light given by the strength of the
absorption lines to differ from the centre of mass. The effects of this
can be represented by allowing for a phase shift in the sine curve, or
more generally by introducing a fictitious eccentricity. They describe a
procedure for detecting any effects of heating on the radial velocity
curve. Firstly one must check for the significance of a fit with an
eccentric orbit. If the fit is not significantly better than a purely
sinusoidal fit, then the semi-amplitude of the curve measured from the
absorption features represents a measure of the true semi-amplitude of
the radial velocity curve of the secondary star, $K_{2}$. If an improved
fit is obtained, this indicates the possible presence of asymmetric heating.
In which case, the data should be treated using a model which includes
the effects of heating.

We found that the fit to the outburst absorption line radial velocity
data with an eccentric orbit is significantly better than that with a
circular orbit. We obtained an eccentricity of $0.119\pm0.023$
(1-$\sigma$ errors) which is significant at the 99 per cent level.
Therefore, the observed value of $K_{obs}$ obtained from a sine wave fit
to the absorption line radial velocity data cannot be taken to represent
the true value of $K_{2}$.

\section{X-ray irradiation of the secondary star}

According to BATSE measurements taken during 1995, the X-ray nova 
GRO J1655--40 continued to have major outburst events in hard X-rays
long after its initial outburst. These include an event seen in late
March of 1995 (Wilson et al.\ 1995), and a further outburst in late
July of 1995 (Harmon et al.\ 1995). The source finally settled into true X-ray
quiescence after  late August of 1995, and was not detected by BATSE for
the remainder of the year (Robinson et al.\ 1996).
The observed X-ray luminosity of X-Ray Nova Sco, as determined from the
BATSE daily averages, varied between 1995 March 18 and March 25 within
the range $5.7\times10^{36} \la L_x \la 2.3\times10^{37}$ \ergss 
(Orosz \& Bailyn 1997).
The outburst radial velocity data was obtained a little over one month later,
during 1995 April and May. 

In order to demonstrate the relative strength of the X-ray heating
in X-Ray Nova Scorpii, we will assume an X-ray luminosity of 
$L_{X}=1.4\times 10^{37}$\ergss, the mean value
of the range quoted above. The intrinsic
luminosity of the secondary, computed from the observations made in the V band
while the system was in quiescence (Orosz \& Bailyn 1997), is approximately
$L_{\mathrm{int}}=1.8\times 10^{35}$\ergss. Given the measured masses and
orbital period of the system (Orosz \& Bailyn 1997), we can use Kepler's 
Third Law to determine the separation, $d$, of the components to be 
approximately 16.8 $R_\odot$. Eggleton's (1983) expression for the 
effective radius of the Roche lobe then determines the radius of the 
secondary, $R_2$, to be about 4.9 $R_\odot$. The \emph{maximum} ratio of
the irradiating flux to the intrinsic flux at the surface of the secondary
(in the limit of normal incidence) is therefore given by
\begin{equation}
\frac{L_x/4 \pi d^2}{L_{\mathrm{int}}/4 \pi R_2^2} \sim 6.6.
\end{equation}

Thus, the incident X-ray flux may exceed the internal flux of the secondary star
by almost a factor of 7, and will result in considerable heating of
the region of the irradiated hemisphere  beyond the 
accretion disc's shadow. Additional evidence for the presence of
irradiation comes from lightcurve fitting performed by Orosz \& Bailyn (1997)
using of optical flux data in the V band taken from 1995 March 18--25 observations. 
The lightcurve exhibits two unequal minima at phases 0 and 0.5, the deepest 
being at phase 0, in contrast with the quiescent lightcurve (Orosz \& Bailyn 1997). 
The shallow minimum at phase 0.5 can be explained by X-ray heating of the
secondary hemisphere facing the compact object. A fitted value of 
$L_x=3.7\times10^{36}$ \ergss is obtained for the X-ray luminosity, slightly
lower than the range quoted above. (However, the discrepancy is not surprising
given the large amount of scatter in the optical lightcurves around phase 0.5). 

We suggest that heating of this magnitude will strongly affect the vertical
temperature gradient in the irradiated atmosphere of the secondary,
and may have significant consequences on the observed
absorption line radial velocity curve. In order to correctly interpret
the radial velocity data taken during outburst, it is therefore
necessary to use a method which directly incorporates these heating
effects.

\section{The X-ray irradiation model}

We modelled the secondary star as a Roche-lobe filling star of mean
effective temperature 6500 K, consistent with its observed spectral type
F3$\sc iv$--F6$\sc iv$. The black hole was assumed to act as a
point-source of radiation, emitting isotropically. The model includes an
accretion disc with opening angle $\beta$, which shadows the region of
the secondary near the inner Lagrangian point from X-rays. The secondary
surface was then divided into grid elements, and an effective temperature
was calculated for each point by combining intrinsic and incident fluxes
(see Shahbaz, Naylor \& Charles 1993; Orosz \& Bailyn 1997; van der Hooft
et al. 1998). We then specify the strength of the absorption lines over the
secondary surface, and integrate to obtain the corresponding mean radial
velocity.

Orosz \& Bailyn (1997) used the absorption lines in the spectral region
5000--6000\AA\ to determine the radial velocity curve of the secondary
star. Using stellar atmospheres (Jacoby, Hunter \& Christian 1984) we
obtained a fit to the total line-flux of the absorption lines in the
spectral region (5000--6000\AA) versus temperature (5000--15000 K)
relationship. The most obvious method would be simply to set the
absorption line strength according to the effective temperature for each
element. However, we must also consider the consequences of
\emph{external} heating. While we expect that the continuum fluxes are
approximated by the star of the correct spectral type, the vertical
temperature gradient in an atmosphere heated internally and externally is
less than the value obtained when heated solely from within. This
produces weaker absorption lines than expected from the effective
temperature. As no satisfactory model exists for the effects of external
heating in stars, we make the following crude approximation (see
Billington, Marsh \& Dhillon 1996). If the incident flux from the X-ray
source exceeds 50 per cent of the unperturbed flux from the secondary,
then we set the line flux for that element to zero; otherwise, the
absorption line strength takes the value corresponding to the effective
temperature of the element, using the stellar atmospheres described
above.

In the case of X-Ray Nova Sco, the incident flux can exceed the intrinsic flux
by almost an order of magnitude, and results in a substantial region of the
secondary having zero absorption strength. Figure 1 shows the
irradiated Roche lobe in the x--z plane, where the z-axis corresponds
to the pole of the secondary and the compact object is at coordinates (1,0).
An X-ray luminosity of $L_{X}=1.4\times 10^{37}$\ergss is assumed,
with a disc angle of $2^\circ$. The shaded areas represent the
regions of the secondary whose absorption line flux has been set to zero 
due to irradiation. The region directly around the inner Lagrangian point  
is shielded by the accretion disc, and hence is unshaded. 

\begin{figure}
\rotate[r]{\epsfxsize=230pt \epsfbox[00 00 700 750]{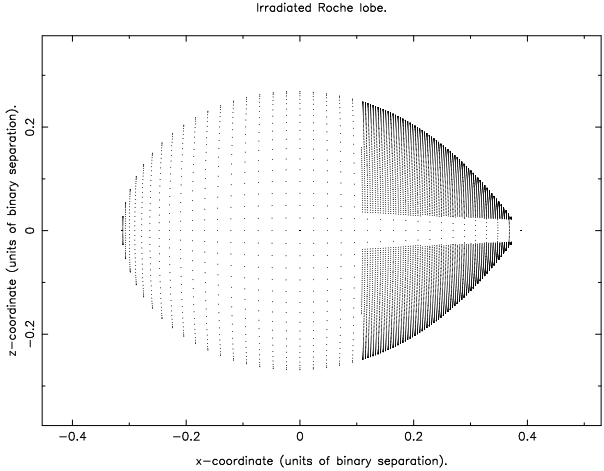}}
\caption{The irradiated Roche lobe in the x--z plane. The compact
object is at coordinates (1,0).
An X-ray luminosity of $L_{X}=1.4\times 10^{37}$\ergss is assumed,
with a disc angle of $2^\circ$. The shaded areas represent the
regions of the secondary whose absorption line flux has been set to zero 
due to irradiation.}
\end{figure}

This decrease in line flux over the irradiated hemisphere of the
secondary leads to significant
asymmetries in the radial velocity curves, and in particular, an increase
in the gradient around phase 0.5. This appears to be consistent with
observations. Being as the incident flux is so far in excess of the
unperturbed flux, the results are not highly sensitive to the exact value
of the X-ray luminosity, nor to the value of the cutoff point (i.e. the
ratio of incident to unperturbed fluxes for which the absorption strength
is set to zero). Although this is clearly a naive and crude model, it
serves to illustrate the extreme effects of X-ray heating, and as we
shall see, it provides a substantially better fit to the data.

Following Wade \& Horne (1988), we may demonstrate the maximum magnitude 
of this effect by considering the extreme
case of a uniform absorption line strength over the back hemisphere of
the secondary and zero absorption over the heated front hemisphere. The
`effective centre' of absorption line strength will then be displaced
from the centre of mass of the star away from the compact object by an
amount $\Delta R_2/R_2=4/3\pi\sim0.42$, where $R_2$ is the radius of the
secondary.  Therefore, the observed amplitude of the radial velocity
variation, $K_{\rm obs}$, will be larger than the true (dynamical)
amplitude, $K_2$, by an amount

\begin{equation}
\Delta K_2=\frac{\Delta R_2}{a_2}K_2=\frac{0.42R_2}{a}(1+q^{-1})K_2,
\end{equation}

\noindent
where $a$ and $a_2$ are the distances from the centre of mass of the
secondary to the centres of mass of the primary and of the system,
respectively. The mass ratio, $q$, is defined as the mass
of the compact object divided by the mass of the secondary star. 
For X-Ray Nova Sco, this leads to a
maximum correction in $K_{\rm obs}$ of around 14 per cent, or $\sim32$
\kms. The mass values derived from the uncorrected $K$ velocity could
thus be in error by as much as $\sim40$ per cent. 

\section{Fitting the outburst radial velocity curve}

Using the model described above we performed a least-squares fit to the
outburst radial velocity data (note that each orbital phase was only
covered once). The free parameters in the model were a
phase shift (phase zero is defined as inferior conjunction of the
secondary star) and the normalisation of the light curve. We performed
least-squares fits to the data using this model, grid searching $K_{2}$
in the range 180--240 \kms, $q$ in the range 2.0--5.0 and $\beta$ in the range
2$^\circ$--14$^\circ$. The effective temperature of 6500 K is appropriate
for an F5$\sc iv$ star, so we used this as the polar temperature. We fixed
the X-ray luminosity at $L_{X}=1.4\times 10^{37}$\ergss, the mean value
of the range quoted by Orosz \& Bailyn (1997), and fitted the outburst
radial velocity curve in the ($K_{2},q,\beta$) coordinate space. 

By fitting the quiescent multi-colour optical light curves of X-Ray Nova Sco 
1994, van der Hooft et al.\ (1998) found a binary inclination in the range
63.7$^\circ$--70.7$^\circ$. Figure 2 shows the \chis fit in the ($K_{2},q$)
plane for these upper and lower inclination limits. The solutions were
obtained by collapsing the minimum \chis solutions along the $\beta$
axis onto the ($K_{2},q$) plane. In effect, we have let $\beta$ run as a
free parameter. We obtained a minimum \rchis of 3.3 at $K_{2}=196$ \kms,
$q=2.8$ and $\beta=2.0^\circ$. The 90 per cent confidence regions are
shown, calculated according to Lampton, Margon \& Bowyer (1976) for 2
parameters, after the error bars had been scaled to give a minimum
\rchis of 1.  (The high value of the minimum \rchis is unsurprising
given the large scatter in the radial velocity data, and suggests that
the error bars have been under-estimated).
Figure 3 (top panel) shows our best fit to the outburst radial velocity
data. A sinusoidal fit is also shown, as was used by Bailyn et al.\
(1995b) and Orosz \& Bailyn (1997). 

We also investigated the effects of changing the level of X-ray
heating on the secondary. The full range of observed X-ray luminosities
was explored: $5.7\times10^{36} \la L_x \la 2.3\times10^{37}$ \ergss,
as determined from the BATSE daily averages (see section 3).
However, it was found that the model was not sensitive to the irradiating
luminosity in this range. For example, decreasing the X-ray luminosity by a factor
of two (from $1.4\times 10^{37}$ to $7\times 10^{36}$\ergss)
only increases the ($K_{2},q$) solutions by 0.6 \kms.

Finally, the effect of a grazing eclipse of the
secondary by the accretion disc was considered in our model. 
A large disc was chosen, of opening angle $14 \degr$
and a radius equal to 80 per cent of the primary's Roche lobe radius,
in order to emphasise the effects on the radial
velocity curve. (Similar disc parameters were used by Orosz \& Bailyn 
(1997) to model the outburst optical lightcurve from March 1995).
Figure 3 (bottom panel, dotted line) shows the residual radial 
velocities obtained by subtracting the sine curve (shown in the top
panel) from a model which contains 
only the eclipse of the secondary by the accretion disc, with no
irradiation effects. Clearly, the residual curve has the right
shape: it is positive just before phase 0.5 and negative just after. 
However, the maximum magnitude (about 13 \kms) is far smaller than
the observed residual for the same phase ($\sim 80$ \kms), and so
the eclipse model provides a totally inadequate fit to the data. 
The residuals obtained from the irradiation model fit, with the
same sine wave subtracted, are also shown (solid line).
The amplitudes are much larger (up to $\sim 40$ \kms), and clearly
give a far better agreement with the data.

\section{The mass of the compact object}

Although we have no reason to doubt the actual values obtained by Orosz
\& Bailyn (1997) for the binary inclination and mass ratio, it should
however be noted that the uncertainties quoted are probably optimistic
given the fact that they have not fully taken into account systematic
effects, which are most definitely present (see their figure 7).
Nevertheless, if we assume $q$ to be 3.0, then this limits $K_{2}$ to
within the range 192--214 \kms\ (90 per cent confidence), which then
constrains the binary mass function to lie in the range 1.93--2.67 \Msun.
Note that this range is \emph{much} lower than that derived by Orosz \&
Bailyn (1997) of $3.24\pm0.09$ \Msun. (We also constrain the systemic
velocity of the binary using the values for the normalisation of the
model fit to the data, obtaining the range -143 to -153 \kms.)
Figure 4 shows the current poorly sampled quiescent radial velocity data of
X-Ray Nova Sco 1994 (Orosz \& Bailyn 1997). We also show the predicted
sinusoidal radial velocity curves for our upper and lower limits on $K_{2}$.
Note that the scatter in the quiescent data exceeds our range in $K_{2}$, and so
cannot be used to restrict acceptable values.

Assuming $q=3.0$, the inclination limits of $63.7^\circ$ and $70.7^\circ$
(van der Hooft et al.\ 1998) and the limits on the mass function obtained
above, we can determine an allowed range for the masses of the black hole 
and the secondary star. We obtain 90 per cent confidence limits of
$4.1<M_{1}<6.6$ \Msun\ and $1.4<M_{2}<2.2$ \Msun\ for the black hole 
and secondary star, respectively.

\section{discussion}

\subsection{The accretion disc opening angle}

The parameter ranges quoted above are derived from our optimum
($K_{2}$,q) solutions shown in figure 2. These are obtained by
collapsing the minimum \chis solutions along the $\beta$ axis. 
Although we could not constrain the disc angle, it should be noted 
that all of these solutions favoured small values
of $\beta$, and our best-fit solution is for $\beta=2\degr$.
Superficially, this appears inconsistent with a 
disc which is transferring enough mass to produce the outburst, and
is below the range of disc angles obtained by several authors for
other X-ray binaries. For example, Mason \& Cordova (1982) analysed
X-ray and optical eclipses of the ADC source 2A 1822--371, from which they deduced
$\beta\sim6$--$14\degr$; Motch et al.\ (1987) estimated $\beta\sim9$--$13\degr$
based on optical observations of 2S 1254--690. However, the above examples
both concern stably accreting systems, whereas X-Ray Nova Scorpii is a
transient. We expect that the disc angle in such a system
may vary dramatically over a dynamical or thermal timescale, which 
is of the order of hours to days for typical disc parameters (Frank,
King \& Raine 1992). Given this variability, a disc angle which is
close to the quiescent value of $\sim2\degr$ (Orosz \& Bailyn 1997),
or at least towards the lower end of the ranges given above,
does not seem unreasonable, despite later observations of the system 
which support larger values (e.g. Hynes et al.\ 1998).

In addition, we must also consider the effects of
irradiation-driven circulation over the surface of the secondary. The
transfer of heated material from the irradiated regions towards the
inner Lagrangian point, and therefore within the disc's shadow, would
produce similar consequences for the radial velocity curve as a
small-angled disc. Furthermore, the obvious asymmetry of the data
around orbital phase 0.5 (when the illuminated hemisphere is directed
towards the line-of-sight) possibly may be explained by 
non-axially symmetric circulation induced by the Coriolis force.
Although a detailed discussion is beyond the scope of this paper, it
has been shown that
such circulation effects are significant. For example, Schandl,
Meyer-Hofmeister \& Meyer (1997) used horizontal heat transfer
in their modelling of the visual
lightcurve of CAL 87; also, the analysis of the optical lightcurve of
HZ Herculis, by Kippenhahn \& Thomas (1979), required circulation
to explain the shape of the lightcurve at minimum.

\subsection{The heliocentric radial velocity of the system} 

Another unique feature in the radial velocity curve of X-Ray Nova Sco 1994 is
the high heliocentric radial velocity of approximately -150 \kms. After
correction for the peculiar motion of the Sun and differential Galactic
rotation, the magnitude of the space velocity of X-Ray Nova Sco 1994 stands out
as being being much higher than any other dynamically identified Galactic
black hole candidate. Brandt, Podsiadlowski \& Sigurdsson (1995) give an
explanation of the high space velocity of X-Ray Nova Sco 1994 in terms of a
delayed black hole creation, which appears to favour the production
of a relatively low black hole mass. In this scenario, the initial collapse
leads to the formation of a neutron star, allowing for a kick normally
associated with a neutron star formation. The neutron star is then
converted into a black hole due either to subsequent accretion of matter
or a phase transition in the compact object. 

According to the
stripped-giant models for the companion star (King 1993; Brandt,
Podsiadlowski \& Sigurdsson 1995) the maximum mass of the secondary is
2.3 \Msun. Our lower limit for the secondary star mass of
$M_{2}>1.4$ \Msun\  implies that a maximum of $\sim0.9$ \Msun\ has
therefore been
available for accretion onto the black hole. Since in the phase
transition scenario, the black hole would initially be formed with a
relatively low mass ($<2$ \Msun, Brown \& Bethe 1994), there is 
insufficient matter available to form the observed lower limit for the
compact object of 4.1 \Msun. The alternative hypothesis in which the black hole 
in X-Ray Nova Sco is formed via an 
intermediate neutron star stage, and then converted into a black hole 
by subsequent accretion of supernova material, therefore appears more 
consistent with our mass limits. 

However, other possible scenarios, such as a prompt black hole formation 
with an associated Blaauw-Boersma kick (see Brandt \& Podsiadlowski 1995), 
cannot be ruled out at this stage.

\section{Conclusions}

We have reanalysed the published outburst absorption line radial velocity
data of X-Ray Nova Sco 1994. We find that as the X-ray source was active
during the observations, one has to model the effects of X-ray
irradiation of the secondary star when interpreting the radial velocity
curve, since the irradiation will affect the strength of the
absorption lines. The observed outburst radial velocity data is fitted
using the X-ray heating model and 90 per cent confidence solutions are
obtained in the ($K_{2},q$) plane. Assuming a binary mass ratio of 3.0 and the
inclination range of $63.7^\circ$ to $70.7^\circ$, we derive limits on
the masses of the binary components: $4.1<M_{1}<6.7$ \Msun\ and
$1.4<M_{2}<2.2$ \Msun\ for the black hole and secondary star,
respectively (90 per cent confidence). This lower limit
for the black hole mass is consistent with the idea that it was formed
as the result of the post-supernova collapse of a neutron star.

We urge future spectroscopic observations of X-Ray Nova Sco 1994 in
\emph{quiescence}, which will enable the
true radial velocity of the secondary star and also the binary mass ratio
to be determined directly. These parameters are crucial in establishing
the true masses of the binary components.

\section*{Acknowledgements}

We would like to thank Jerry Orosz for providing the radial velocity
data.

\begin{figure*} 
\rotate[l]{\epsfxsize=500pt \epsfbox[00 00 700 750]{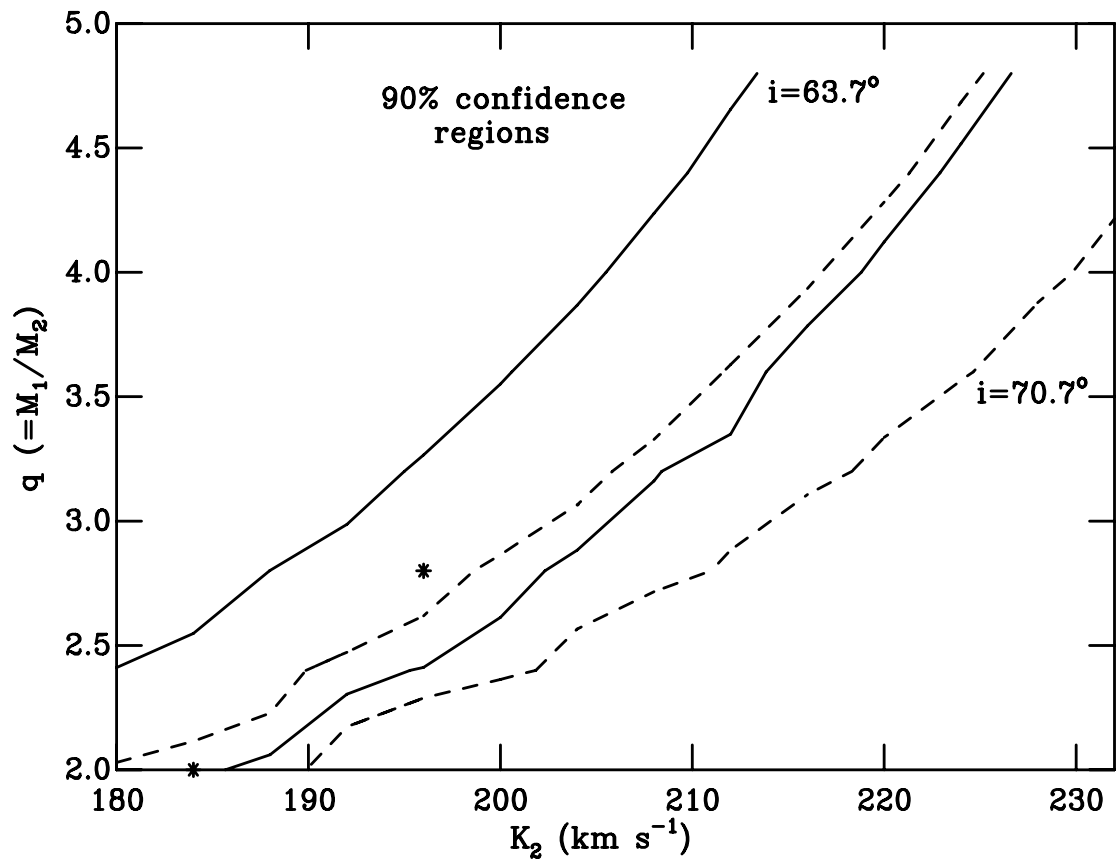}}
\caption{
The 90 per cent confidence level solutions for model fits to the
outburst radial velocity data of X-Ray Nova Sco 1994 are shown. An X-ray
luminosity of $L_{X}=1.4\times 10^{37}$\ergss\ was used. The ($K_{2},q$)
solutions were obtained by collapsing the minimum \chis solutions along
the $\beta$ axis. The regions bounded by the solid and dashed lines
contain fits using $i=63.7^\circ$ and $70.7^\circ$, respectively. The
stars show the best fit solutions for the two inclination limits.}
\end{figure*}

\begin{figure*} 
\rotate[l]{\epsfxsize=500pt \epsfbox[00 00 700 750]{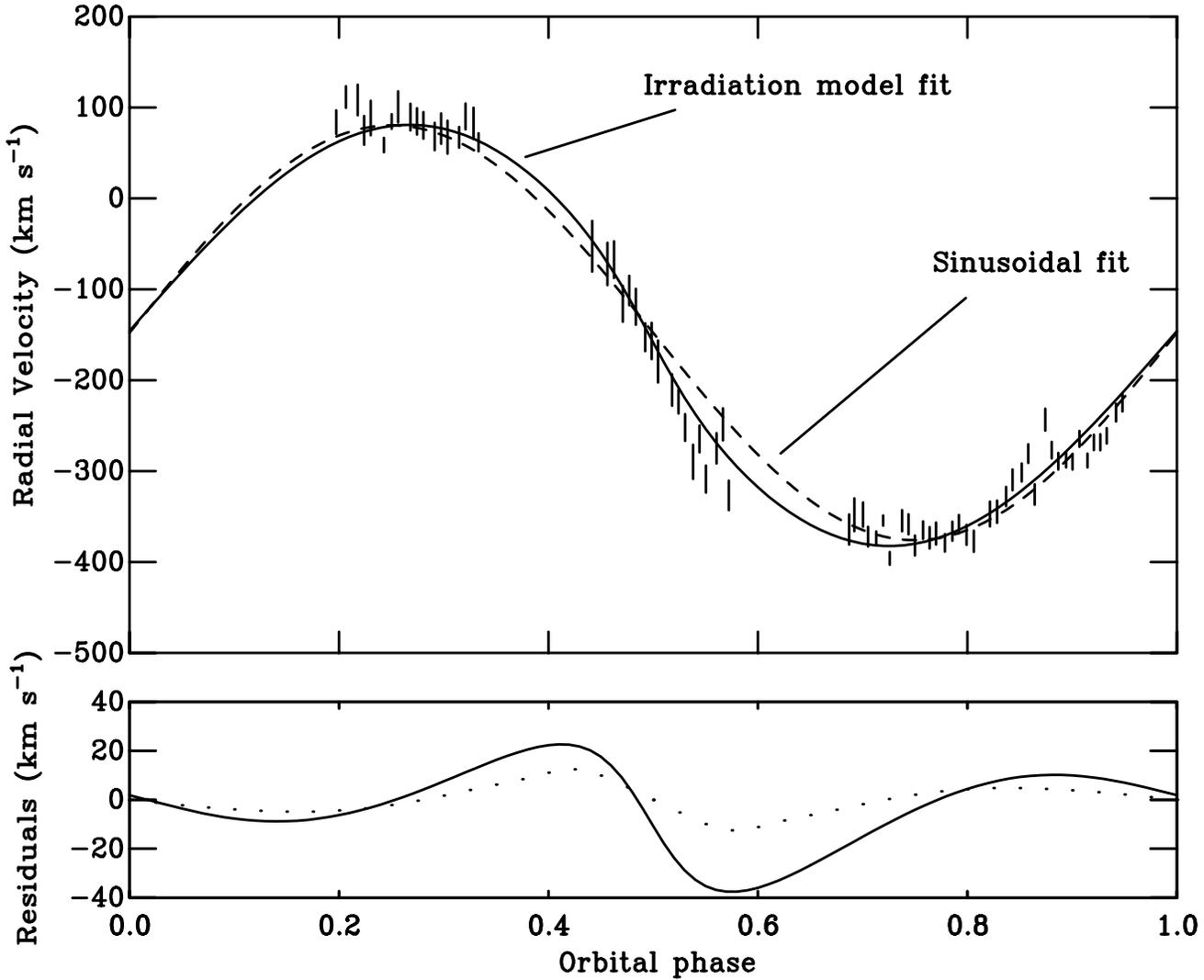}}
\caption{
The outburst radial velocity curve of X-Ray Nova Sco 1994 obtained from the
absorption lines of the secondary star (Orosz \& Bailyn 1997).
\emph{Top panel}: The data is
fitted with a model which includes the effects of X-ray irradiation of
the secondary star. The solid line shows the best model fit using
$L_{X}=1.4\times 10^{37}$\ergss, $i=63.7^\circ$, $\beta=2^\circ$,
$q=2.8$ and $K_{2}=196$ \kms. The dashed line is a sinusoidal fit to the
data.
\emph{Bottom panel}: The solid line shows the residual radial 
velocity obtained by subtracting the sine curve (shown in the top
panel) from the irradiation model fit. 
The dotted line is the residual using a model which contains 
only the eclipse of the secondary by the accretion disc, with no
irradiation effects, and again with the sine curve subtracted. }
\end{figure*}

\begin{figure*} 
\rotate[l]{\epsfxsize=500pt \epsfbox[00 00 700 750]{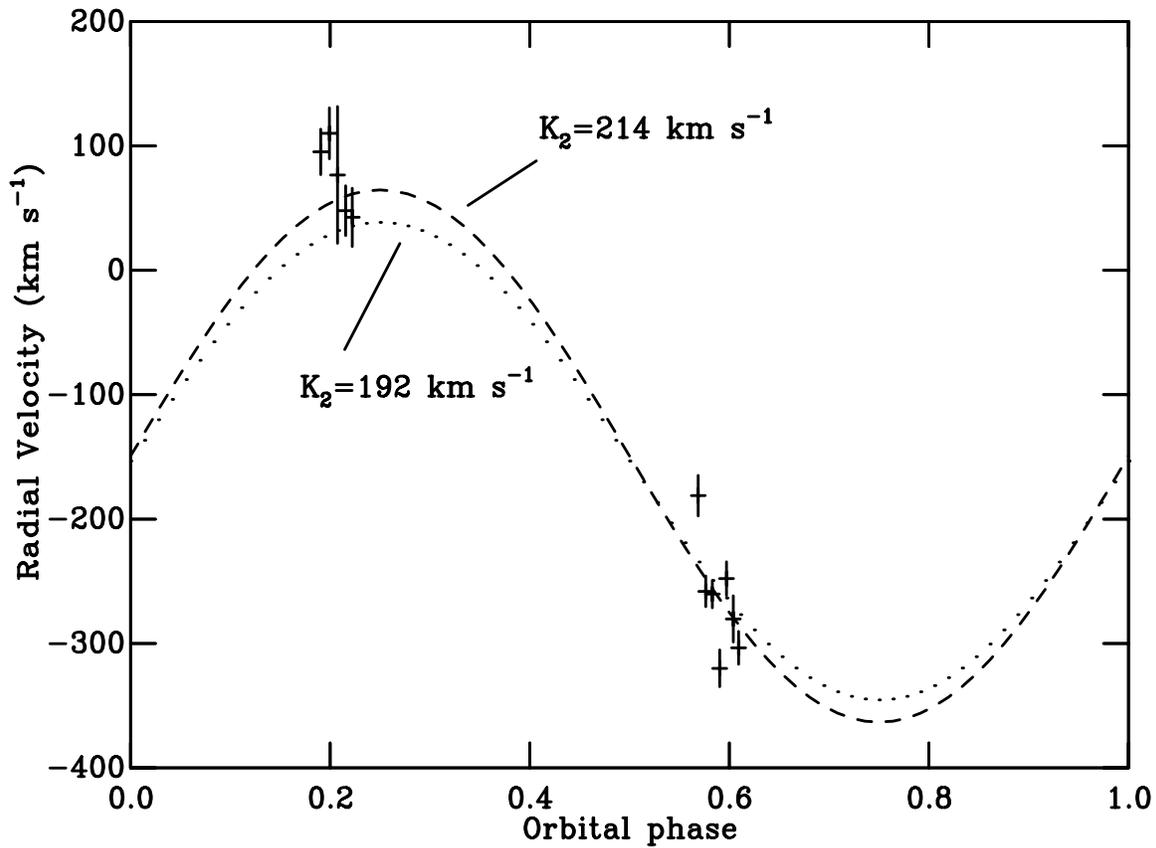}}
\caption{
The poorly sampled quiescent radial velocity data of X-Ray Nova Sco 1994 (Orosz
\& Bailyn 1997). The solid and dotted lines show the predicted
sinusoidal radial velocity curves of the secondary star during
quiescence, with $K_{2}=192$ \kms\ and 214 \kms, respectively. These
limits were obtained using our ($K_{2},q$) fits and assuming $q=3.0$ (see
section 5). }
\end{figure*}

\end{document}